\documentclass[runningheads,final]{llncs}

\usepackage{ifluatex}

\usepackage{comment}

\usepackage{graphicx}

\usepackage{float}

\usepackage{hyperref}
\hypersetup{
  unicode,
  colorlinks = true,
  allcolors = black,
  urlcolor = blue,
  pdfborder = 0 0 0,
}

\ifluatex
\usepackage[newfloat,draft=false,chapter,finalizecache]{minted}
\else
\usepackage[newfloat,draft=false,chapter,frozencache]{minted}
\fi
\newmintinline{c}{style=bw}

\usepackage{multirow}

\usepackage{subcaption}

\usepackage{longtable}

\usepackage[nameinlink,capitalize]{cleveref}
\crefname{sublisting}{Listing}{Listings}
\Crefname{sublisting}{Listing}{Listings}

\usepackage{minibox}

\ifluatex
\usepackage{tikz}
\usetikzlibrary{calc}
\usetikzlibrary{decorations}
\usetikzlibrary{decorations.pathreplacing}
\usetikzlibrary{decorations.text}
\usetikzlibrary{decorations.pathmorphing}
\usetikzlibrary{decorations.markings}
\usetikzlibrary{fit}
\usetikzlibrary{arrows}    % Deprecated, use arrows.meta
\usetikzlibrary{shapes}
\usetikzlibrary{shadows}
\usetikzlibrary{automata}
\usetikzlibrary{positioning}
\usetikzlibrary{petri}
\usetikzlibrary{chains}
\usetikzlibrary{fadings}
\usetikzlibrary{matrix}
\usetikzlibrary{patterns}
\usetikzlibrary{mindmap}
\usetikzlibrary{graphs}			% TikZ 3.0.0
\usetikzlibrary{graphdrawing}	% TikZ 3.0.0
\usetikzlibrary{quotes}			% TikZ 3.0.0
\usetikzlibrary{babel}          % Vor polyglossia
\usetikzlibrary{hobby}
\usetikzlibrary{arrows.meta}
\usetikzlibrary{backgrounds}
\usetikzlibrary{bending}
\usetikzlibrary{external}

\usegdlibrary{trees}			% TikZ 3.0.0
\usegdlibrary{layered}			% TikZ 3.0.0
\usegdlibrary{force}			% TikZ 3.0.0
\usegdlibrary{circular}			% TikZ 3.0.0
\usegdlibrary{routing}			% TikZ 3.0.0

\usepackage{pgfplots}
\pgfplotsset{compat=1.12}

\makeatletter
\def\myfoldheight{0.5}

\def\myshapepath{
\pgfextract@process\northwest{
        \southwest\pgf@xa=\pgf@x
        \northeast
        \pgf@x=\pgf@xa
}

\pgfextract@process\southeast{
        \southwest\pgf@ya=\pgf@y
        \northeast
        \pgf@y=\pgf@ya
}

\pgfextract@process\northfold{
        \pgfpointdiff{\southwest}{\northeast}
        \northeast
        \advance\pgf@x-\myfoldheight\pgf@y
}

\pgfextract@process\eastfold{
        \pgfpointdiff{\southwest}{\northeast}
         \northeast
        \advance\pgf@y-\myfoldheight\pgf@y
}

\pgfextract@process\fold{
        \northfold\pgf@xa=\pgf@x
        \eastfold
        \pgf@x=\pgf@xa
}

        \pgfpathmoveto{\southwest}
        \pgfpathlineto{\northwest}
        \pgfpathlineto{\northfold}
        \pgfpathlineto{\eastfold}
        \pgfpathlineto{\southeast}
        \pgfpathclose
}

\def\myshapeanchorborder#1#2{
    \pgftransformreset % without this, the intersection commands yield strange results
    \pgf@relevantforpicturesizefalse % don't include drawings in bounding box
    \pgfintersectionofpaths{
        \myshapepath
    }{
        \pgfpathmoveto{
            \pgfpointadd{
                \pgfpointdiff{\southwest}{\northeast}\pgf@xc=\pgf@x \advance\pgf@xc by \pgf@y % calculate a distance that is guaranteed to be outside the shape
                \pgfpointscale{
                    \pgf@xc
                }{
                    \pgfpointnormalised{
                        #2
                    }
                }
            } {
                #1
            }
        }
        \pgfpathlineto{#1}
    }
    \pgfpointintersectionsolution{1}
}

\newdimen\myshapedimenx
\newdimen\myshapedimeny

\pgfdeclareshape{file}{
    \inheritsavedanchors[from=rectangle]
    \inheritanchor[from=rectangle]{center}
    \inheritanchor[from=rectangle]{mid}
    \inheritanchor[from=rectangle]{base}

    \inheritanchor[from=rectangle]{west}
    \inheritanchor[from=rectangle]{east}
    \inheritanchor[from=rectangle]{north}
    \inheritanchor[from=rectangle]{south}

    \inheritanchor[from=rectangle]{south west}
    \inheritanchor[from=rectangle]{south east}
    \inheritanchor[from=rectangle]{north west}
    \inheritanchor[from=rectangle]{mid west}
    \inheritanchor[from=rectangle]{mid east}
    \inheritanchor[from=rectangle]{base west}
    \inheritanchor[from=rectangle]{base east}

    \backgroundpath{
       \myshapepath
     }

    \foregroundpath{
     \pgfpathmoveto{\northfold}
     \pgfpathlineto{\fold}
     \pgfpathlineto{\eastfold}
    }

  \anchorborder{%
    \pgf@xb=\pgf@x% xb/yb is target
    \pgf@yb=\pgf@y%
    \southwest%
    \pgf@xa=\pgf@x% xa/ya is se
    \pgf@ya=\pgf@y%
    \northeast%
    \advance\pgf@x by-\pgf@xa%
    \advance\pgf@y by-\pgf@ya%
    \pgf@xc=.5\pgf@x% x/y is half width/height
    \pgf@yc=.5\pgf@y%
    \advance\pgf@xa by\pgf@xc% xa/ya becomes center
    \advance\pgf@ya by\pgf@yc%
    \edef\pgf@marshal{%
      \noexpand\pgfpointborderrectangle
      {\noexpand\pgfqpoint{\the\pgf@xb}{\the\pgf@yb}}
      {\noexpand\pgfqpoint{\the\pgf@xc}{\the\pgf@yc}}%
    }%
    \pgf@process{\pgf@marshal}%
    \advance\pgf@x by\pgf@xa%
    \advance\pgf@y by\pgf@ya%
  }

}

\makeatother

\usepackage[textsize=miniscule,obeyDraft]{todonotes}
\makeatletter
\newcommand\miniscule{\@setfontsize\miniscule{4}{5}}
\makeatother
\newcommand\meinersbur[2][]{\todo[%
linecolor=yellow!50!black,backgroundcolor=yellow,bordercolor=yellow!50!black,#1]{#2}}
\newcommand\hfinkel[2][]{\todo[%
linecolor=blue,backgroundcolor=blue!30!white,bordercolor=blue,#1]{#2}}

\else
\usepackage{tikzexternal}
\newcommand\meinersbur[2][]{}
\newcommand\hfinkel[2][]{}
\fi

\let\svthefootnote\thefootnote
\newcommand\blankfootnote[1]{%
  \let\thefootnote\relax\footnotetext{#1}%
  \let\thefootnote\svthefootnote%
}

\usepackage{pgfmath}

\makeatletter
\newsavebox{\my@scalepar@TempBox}
\newenvironment{scalepar}[1]{%
\def\my@DoScale{\scalebox{#1}}%
\begin{lrbox}{\my@scalepar@TempBox}%
\pgfmathparse{\textwidth/#1}%
\begin{minipage}{\pgfmathresult pt}%
}{%
\end{minipage}%
\end{lrbox}%
\my@DoScale{\usebox{\my@scalepar@TempBox}}%
}
\makeatother

\begin{document}

\title{A Proposal for Loop-Transformation Pragmas}
%\titlerunning{Loop-Transformation Pragmas}

\author{Michael Kruse, Hal Finkel}

\institute{Argonne Leadership Computing Facility,\\
Argonne National Laboratory, Argonne IL 60439, USA\\
\email{mkruse@anl.gov, hfinkel@anl.gov}}

\maketitle

\begin{abstract}
Pragmas for loop transformations, such as unrolling, are implemented in most mainstream compilers. 
They are used by application programmers because of their ease of use compared to directly modifying the source code of the relevant loops.
We propose additional pragmas for common loop transformations that go far beyond the transformations today's compilers provide and should make most source rewriting for the sake of loop optimization unnecessary.
To encourage compilers to implement these pragmas, and to avoid a diversity of incompatible syntaxes, we would like to spark a discussion about an inclusion to the OpenMP standard.

\end{abstract}

% !TeX encoding = UTF-8
% !TeX spellcheck = en_US
% !TeX root = paper.tex

\newcommand\pragma[1]{\texttt{\#pragma~#1}}
\newcommand\pragmaomp[1]{\texttt{\#pragma~omp~#1}}
\newcommand*\joop[1]{\texttt{#1}}
\newcommand*\data[1]{\texttt{#1}}
\newcommand*\sect[1]{\texttt{#1}}
\newcommand\syntax[1]{\textbf{\texttt{#1}}}
\newcommand\placeholder[1]{\textrm{\textit{#1}}}

%%%%%%%%%%%%%%%%%%%%%%%%%%%%%%%%%%%%%%%%%%%%%%%%%%%%%%%%%%%%%%%%%%%%%%%%%%%%%%%%%
%%%%%%%%%%%%%%%%%%%%%%%%%%%%%%%%%%%%%%%%%%%%%%%%%%%%%%%%%%%%%%%%%%%%%%%%%%%%%%%%%
\section{Motivation}\label{sct:motivation}

Almost all processor time is spent in some kind of loop, and as a result, loops are a primary targets for program-optimization efforts.
One method for optimizing loops is annotating them with OpenMP pragmas, such as \pragmaomp{parallel~for}, which executes the loop iterations in multiple threads, or \pragmaomp{simd} which instructs the compiler to generate vector instructions.

% Advantages over manual transformation
Compared to manually parallelizing the relevant code (e.g. using the pthreads library) or manually vectorizing the relevant code (e.g. using SIMD-intrinsics or assembly), annotating a loop yields much higher programmer productivity.
In conjunction with keeping the known-to-work statements themselves unchanged, we can expect less time spent on optimizing code and fewer bugs.
Moreover, the same code can be used for multiple architectures that require different optimization parameters, and the impact of adding an annotation can be evaluated easily.
Directly applied transformations also make the source code harder to understand since most of the source lines will be for the sake of performance instead of the semantics.

% Separating semantics from operations
Pragmas allow separating the semantics-defining code and the performance-relevant directives.
Using \texttt{\_Pragma("...")}, the directives do not necessarily need appear adjacent to the loops in the source code, but can, for instance, be \texttt{\#include}d from another file.
%Frameworks such as Halide~\cite{halide} and Tensor Comprehensions~\cite{tensorcomprehensions} follow this approach, albeit in the form of a library.

Most compilers implement additional, but compiler-specific pragmas.
Often these have been implemented to give the programmer more control over its optimization passes, but without a systematic approach for loop transformations.  
\Cref{tbl:proprietarypragmas} shows a selection of pragmas supported by popular compilers.

{\small%
\begin{longtable}{lll}
	Transformation                                            & Syntax Example                                               & Compiler                                                                            \\ \hline
\endhead
	Threading                                                 & \pragmaomp{parallel for}                                     & OpenMP~\cite{openmp}                                                                \\
	                                                          & \pragma{loop(hint\_parallel(0)}                              & msvc~\cite{msvcloop}                                                                \\
	                                                          & \pragma{parallel}                                            & icc~\cite{iccmanual}                                                                \\
	Unrolling                                                 & \pragma{unroll}                                              & \minibox{icc~\cite{iccmanual}, xlc~\cite{xlcmanual}, clang~\cite{clangattributes}}  \\
	                                                          & \pragma{clang loop unroll(enable)}                           & clang~\cite{clangextensions}                                                        \\
	                                                          & \pragma{GCC unroll \placeholder{n}}                          & gcc~\cite{gccpragmas}                                                               \\
	Unroll and jam                                            & \pragma{unroll\_and\_jam}                                    & icc~\cite{iccmanual}                                                                \\
	                                                          & \pragma{unrollandfuse}                                       & xlc~\cite{xlcmanual}                                                                \\
	                                                          & \pragma{stream\_unroll}                                      & xlc~\cite{xlcmanual}                                                                \\
	Loop fusion                                               & \pragma{nofusion}                                            & icc~\cite{iccmanual}                                                                \\
	Loop distribution                                         & \pragma{distribute\_point}                                   & icc~\cite{iccmanual}                                                                \\
	                                                          & \pragma{clang loop distribute(enable)}                       & clang\cite{clangextensions}                                                         \\
	Loop blocking                                             & \pragma{block\_loop(\placeholder{n},\placeholder{loopname})} & xlc~\cite{xlcmanual}                                                                \\
	Vectorization                                             & \pragmaomp{simd}                                             & OpenMP~\cite{openmp}                                                                \\
	                                                          & \pragma{simd}                                                & icc~\cite{iccmanual}                                                                \\
	                                                          & \pragma{vector}                                              & icc~\cite{iccmanual}                                                                \\
	                                                          & \pragma{loop(no\_vector)}                                    & msvc~\cite{msvcloop}                                                                \\
	                                                          & \pragma{clang loop vectorize(enable)}                        & clang~\cite{clangextensions,clangvectorizers}                                       \\
	Interleaving                                              & \pragma{clang loop interleave(enable)}                       & clang~\cite{clangextensions,clangvectorizers}                                       \\
	Software pipelining                                       & \pragma{swp}                                                 & icc~\cite{iccmanual}                                                                \\
	Offloading                                                & \pragmaomp{target}                                           & OpenMP~\cite{openmp}                                                                \\
	                                                          & \pragma{acc kernels}                                         & OpenACC~\cite{openacc}                                                              \\
	                                                          & \pragma{offload}                                             & icc~\cite{iccmanual}                                                                \\
	\multirow{2}{*}{\minibox{Assume iteration\\independence}} & \pragma{pragma ivdep}                                        & icc~\cite{iccmanual}                                                                \\
	                                                          & \pragma{GCC ivdep}                                           & gcc~\cite{gccpragmas}                                                               \\
	                                                          & \pragma{loop(ivdep)}                                         & msvc~\cite{msvcloop}                                                                \\
	Iteration count                                           & \pragma{loop\_count(\placeholder{n})}                        & icc~\cite{iccmanual}                                                                \\
	Loop naming                                               & \pragma{loopid(\placeholder{loopname})}                      & xlc~\cite{xlcmanual}\\
	\caption{Loop pragmas and the compilers which support them}\label{tbl:proprietarypragmas}
	% offload: OpenACC, CUDA
\end{longtable}}

\begin{comment}
% Example
For an illustration of how such loop transformation pragmas can be used to construct an optimized source base, see \cref{lst:unrolling}.
The first listing shows the original code as written by a programmer.
Loop unrolling with an unroll-factor of four has been applied to the code in \cref{lst:unrolled}.
In contrast, \cref{lst:pragmaunroll} should yield the same output while preserving the readability.

\begin{listing}
\begin{subfigure}[b]{.5\linewidth}
\begin{minted}{c}
for (int i=0; i<n; i+=1)
  Statement(i);
\end{minted}
\vspace*{-3ex}%
\caption{A simple loop}\label{lst:simpleloop}

\begin{minted}{c}
#pragma unroll 4
for (int i=0; i<n; i+=1)
  Statement(i);
\end{minted}
\vspace{-3ex}%
\caption{Loop with an unroll annotation}\label{lst:pragmaunroll}
\end{subfigure}%
\begin{subfigure}[b]{.5\linewidth}
\begin{minted}{c}
  int i = 0;
  for (; i+3 < n; i+=4) {
    Statement(i);
    Statement(i+1);
    Statement(i+2);
    Statement(i+3);
  }
  if (i >= n) goto done;
  Statement(i);
  if (i+1 >= n) goto done;
  Statement(i+1);
  if (i+2 >= n) goto done;
  Statement(i+2);
done:
\end{minted}
% if (i   < n) Statement(i);
% if (i+1 < n) Statement(i+1);
% if (i+2 < n) Statement(i+2);
\vspace*{-3ex}%
\caption{Loop after manual unrolling}\label{lst:unrolled}
\end{subfigure}%
\vspace*{-2ex}%
\caption{Loop unrolling example}\label{lst:unrolling}
\end{listing}
\end{comment}

% Large example
Our vision is to enable optimizations such as in \cref{lst:gemm}, which separates the algorithm from its optimization thus keeping the code readable and maintainable.
It also illustrates how different transformations can be applied for different compilation targets.
Many BLAS implementations of dgemm use this pattern~\cite{low16}, but pre-transformed in the source.
As demonstrated in \cref{evaluation}, the performance is comparable to that of optimized BLAS libraries.

\begin{listing}
\begin{minted}{c}
for (int i = 0; i < M; i+=1)
  for (int j = 0; j < N; j+=1) {
    #pragma omp section id(zero)
    { C[i][j] = 0; }
    for (int k = 0; k < K; k+=1)
      C[i][j] += A[i][k] * B[k][j];
  }

#if __haswell__
  #pragma omp loop(i,j) distribute sections(zero,k)
  #pragma omp loop(i,j,k) tile sizes(96,2048,256) \
      pit_ids(i1,j1,k1) tile_ids(i2,j2,k2)
  #pragma omp loop(i1,...,j2) interchange permutation(j1,k1,i1,j2,i2)
  #pragma omp loop(i2,k2) pack array(A)
  #pragma omp loop(j2,k2) pack array(B)
  #pragma omp loop(i2) simd
#elif __skylake__
  [...]
\end{minted}
\vspace*{-4ex}%
\caption{Optimization of matrix-matrix multiplication using our proposed pragmas.
The tile sizes were derived using the analytical model in \cite{low16} for Intel's Kaby Lake architecture.}
\label{lst:gemm}
\end{listing}

\begin{comment}
\begin{listing}
\begin{minted}[fontsize=\small]{c}
for (int i = 0; i < M; i+=1)
  for (int j = 0; j < N; j+=1)
    C[i][j] = 0;
for (int j1 = 0; j1 < N; j1+=2048)
  for (int k1 = 0; k1 < K; k1+=256) {
    double Bpacked[256][2048];
    for (int k2 = k1; k2 < K && k2 < k1+256; k2+=1)
      for (int j2 = 0; j2 < N && j2 < j1+2048; j2+=1)
        Bpacked[k2-k1][j2-j1] = B[k2][i2];
    for (int i1 = 0; i1 < M; i1+=96) {
      double Apacked[96][256];
      for (int i2 = i1; i2 < M && i2 < i1+96; i2+=1)
        for (int k2 = k1; k2 < K && k2 < k1+256; k2+=1)
          Apacked[i2-i1][k2-k1] = A[i2][k2];
      for (int j2 = 0; j2 < N && j2 < j1+2048; j2+=1)
        #pragma omp simd
        for (int i2 = i1; i2 < M && i2 < i1+96; i2+=1)
          for (int k2 = k1; k2 < K && k2 < k1+256; k2+=1)
            C[i2][j2] += Apacked[i2-i1][k2-k1] * Bpacked[k2-k1][j2-j1];
      }
  }
\end{minted}
\vspace*{-4ex}%
\caption{Approximately equivalent code for the matrix-matrix multiplication in \cref{lst:gemm}.}
\label{lst:gemmexpanded}
\end{listing}
\end{comment}

% Autotuning
Many existing loop annotation schemes, including OpenMP, require the user to guarantee some ``safety'' conditions (i.e., that the loop is safe to parallelize as specified) for the use of the annotations to be valid. 
%Requiring the user to guarantee the validity of loop transformations is not always practical.
Making the user responsibility for the validity of a loop transformation may not always be practical, for instance if the code base is large or the loop is complex. 
Compiler assistance, e.g. providing warnings that a transformation might not be safe, can be a great help.
Our use case is an autotuning optimizer, which by itself has only a minimal understanding of the code semantics. 
Thus, we propose a scheme whereby the compiler can ensure that the semantics of the program remains the same (by ignoring the directive or exiting with an error) when automatically-derived validity conditions are unsatisfied.

%%%%%%%%%%%%%%%%%%%%%%%%%%%%%%%%%%%%%%%%%%%%%%%%%%%%%%%%%%%%%%%%%%%%%%%%%%%%%%%%%
%%%%%%%%%%%%%%%%%%%%%%%%%%%%%%%%%%%%%%%%%%%%%%%%%%%%%%%%%%%%%%%%%%%%%%%%%%%%%%%%%
\section{Proposal}

In this publication, we would like to suggest and discuss the following ideas as extensions to OpenMP:

\begin{enumerate}
\item The possibility to assign identifiers to loops and code sections, and to refer to them in loop transformations
\item A set of new loop transformations
\item A common syntax for loop transformation directives
%\item A proof-of-concept implementation in Clang using Polly~\cite{polly} as loop transformer
\end{enumerate}

In the following, we discuss these elements of the proposal and a proof-of-concept implementation in Clang.
We do not discuss syntax or semantics specific to Fortran, but the proposal should be straightforward to adapt to cover Fortran.

%%%%%%%%%%%%%%%%%%%%%%%%%%%%%%%%%%%%%%%%%%%%%%%%%%%%%%%%%%%%%%%%%%%%%%%%%%%%%%%%%
\subsection{Composition of Transformations}\label{sct:loopnaming}

% Stacking transformations
As loop transformations become more complex, one may want to apply more than one transformation on a loop, including applying the same transformation multiple times.
For transformations that result in a single loop this can be accomplished by ``stacking up'' transformations.
Like a regular pragma that applies to the loop that follows, such a transformation directive can also be defined to apply to the output of the following transformation pragma.

The order of transformations is significant as shown in \cref{lst:compositionexample}.
In the former the order of execution will be the exact reversal of the original loop, while in the latter, groups of two statements keep their original order.

\begin{listing}
\begin{subfigure}[b]{.5\linewidth}
\begin{minted}{c}
#pragma omp unroll factor(2)
#pragma omp reverse
for (int i=0; i<n; i+=1)
  Statement(i);
\end{minted}
\vspace*{-3ex}%
\caption{Reversal followed by partial unrolling}\label{lst:unrollreverse}
\end{subfigure}%
\begin{subfigure}[b]{.5\linewidth}
\begin{minted}{c}
#pragma omp reverse
#pragma omp unroll factor(2)
for (int i=0; i<n; i+=1)
  Statement(i);
\end{minted}
\vspace*{-3ex}%
\caption{Partial unrolling followed by reversal}\label{lst:reverseunroll}
\end{subfigure}%
\vspace*{-2ex}%
\caption{Transformation composition of loop unrolling and loop reversal}\label{lst:compositionexample}
\end{listing}

%%%%%%%%%%%%%%%%%%%%%%%%%%%%%%%%%%%%%%%%%%%%%%%%%%%%%%%%%%%%%%%%%%%%%%%%%%%%%%%%%
\subsection{Loop/Section Naming}

In case a transformation has more than one input- or output-loop, transformations or its follow-up transformations require a means to identify which loop to apply to.
%Two examples are loop distribution and strip-mining.

As a solution, we allow assigning names to loops and refer to them in other transformations.
An existing loop can be given a name using
\begin{minted}[escapeinside=!!]{c}
#pragma omp id(!\placeholder{loopname}!)
for (int i=0; i<n; i+=1) ...
\end{minted}
and loops from transformations get their identifier as defined by the transformation, for instance as an option. 
As an example, \cref{lst:nameddist} shows a loop that is strip-mined.
The inner loop is vectorized while the outer is parallelized.
\begin{comment}
It should have the same effect as
\begin{minted}{c}
#pragma omp parallel for schedule(static) simd
for (int i = 0; i < n; i+=1)
  Statement(i);
\end{minted}
with an implied SIMD-width of 4.
\end{comment}

\begin{listing}
\begin{minted}{c}
#pragma omp loop(outer) thread_parallelize
#pragma omp loop(inner) vectorize
#pragma omp stripmine strip_width(4) strip_id(inner) pit_id(outer)
for (int i = 0; i < n; i+=1)
  Statement(i);
\end{minted}
\vspace*{-4ex}%
\caption{Loop strip mining with follow-up transformations}\label{lst:nameddist}%
\end{listing}

% Section names
In some cases, it may be necessary to not only assign an identifier to the whole loop, but also to individual parts of the loop body.
An example is loop distribution: The content of the new loops must be defined and named.
Like for (canonical) for-loops, OpenMP also has a notion of sequential code pieces: sections and tasks.
We reuse this idea to also assign names to parts of a loop, as shown in \cref{lst:distributionexample}.
The transformation results in two loops, named \texttt{loopA} and \texttt{loopB}, containing a call to \texttt{StatementA}, respectively \texttt{StatementB}.

\begin{listing}
\begin{minted}{c}
#pragma omp section(A,B) distribute distributed_ids(loopA, loopB)
for (int i=0; i<n; i+=1) {
  #pragma omp id(A)
  { StatementA(i); }
  #pragma omp id(B)
  { StatementB(i); }
}
\end{minted}
\vspace*{-4ex}%
\caption{Loop distribution example}\label{lst:distributionexample}
\end{listing}

Loop and section names form a common namespace, i.e. it is invalid to have a section and a loop with the same name.
When being used in a loop transformation, a section name stands for the loop that forms when distributed from the remainder of the loop body.

\begin{comment}
\subsubsection{Pragma Location}

With the loop being specified explicitly in the directive, the location of the pragma is of less significance.
We have multiple options here:

\begin{enumerate}
\item Ignore the location of pragma, it has no significance...
	\begin{enumerate}
	\item as long as it is in the same function.
	\item and may appear anywhere in the file.
		The creates scoping problems in that multiple loops with the same name may occur in some function.
	\end{enumerate}
\item The pragma may only appear in front of the first or topmost loop that is involved.
	If it occurs anywhere, the compiler should emit an error.
\item Assign additional semantics to the location of the pragma: After applying the transformation, \emph{move} the loop nest to where the pragma. occurs.
\end{enumerate}

We suggest to use option 2) and relax the requirement an argument for one of the other options is found.
\end{comment}

% Implicit names
To avoid boilerplate pragmas to assign loop names, loops are assigned implicit names.
Every loop is assigned the name of its loop counter variable, unless:
\begin{itemize}
\item it has a \pragma{loop id(..)} annotation,
\item some other loop has an annotation with that name,
\item or there is another loop counter variable with the same name.
\end{itemize}

For instance, fusing the loops in \cref{lst:forcounterimplicitnames} is legal, but the compiler should report an ambiguity in \cref{lst:ambiguousname}.

\begin{listing}
\begin{subfigure}[b]{0.5\linewidth}
\begin{minted}{c}
#pragma omp loop(i,j) fuse
for (int i=0; i<n; i+=1) { .. }
for (int j=0; j<n; j+=1) { .. }
\end{minted}
\vspace*{-3ex}%
\caption{Working example}\label{lst:forcounterimplicitnames}
\end{subfigure}%
\begin{subfigure}[b]{0.5\linewidth}
\begin{minted}{c}
#pragma omp loop(i,i) fuse
for (int i=0; i<n; i+=1)  { .. }
for (int i=0; i<n; i+=1)  { .. }
\end{minted}
\vspace*{-3ex}%
\caption{Ambiguous implicit name}\label{lst:ambiguousname}
\end{subfigure}
\vspace*{-4ex}%
\caption{Implicit loop name example}
\end{listing}

%%%%%%%%%%%%%%%%%%%%%%%%%%%%%%%%%%%%%%%%%%%%%%%%%%%%%%%%%%%%%%%%%%%%%%%%%%%%%%%%%
\subsection{Transformations}\label{sct:transformations}

In addition to the loop transformations mentioned in \cref{tbl:proprietarypragmas}, there are many more transformations compilers could implement.
\Cref{tbl:loopdirectives,tbl:codedirectives} contain some pragmas that could be supported.

% Loop transformations
\Cref{tbl:loopdirectives} contains the directives that apply to loops, many of which change the order of execution.
Transformations such as unrolling and unswitching do not affect the code semantics and therefore can always be applied.
The additional assume- and expect-directives do not transform code, but give hints to the compiler about intended semantic properties.

{\small%
\begin{longtable}{ll}
	Directive                                               & Short description                                    \\ \hline
	\endhead
%		\joop{id}                                   & Assign an unique name                                \\
	\joop{parallel for}                                     & OpenMP thread-parallelism                            \\
	\joop{simd}                                             & Vectorization                                        \\
	\joop{unroll}                                           & Loop unrolling                                       \\
	\joop{split}                                            & Index set splitting                                  \\
	\joop{peel}                                             & Loop peeling (special kind of index set splitting)   \\
	\joop{specialize}                                       & Loop versioning                                      \\
	\joop{unswitch}                                         & Loop unswitching (special kind of loop versioning)   \\
	\joop{shift}                                            & Add offset to loop counter                           \\
	\joop{scale}                                            & Multiply loop counter by constant                    \\
	\joop{coalesce}                                         & Combine nested loops into one                        \\
	\joop{concatenate}                                      & Combine sequential loops into one                    \\
	\joop{interchange}                                      & Permute order of nested loops                        \\
	\joop{stripmine}                                        & Strip-mining                                         \\
	\joop{block}                                            & Like strip-mining, but with constant sized outer loop \\
	\joop{tile}                                             & Tiling (combination of strip-mining and interchange) \\
%	\joop{decompose}                                        & Generalized strip-mining, blocking and tiling         \\
	\joop{reverse}                                          & Inverse iteration order                              \\
	\joop{distribute}                                       & Split loop body into multiple loops                  \\
	\joop{fuse}                                             & Loop Fusion/Merge: Inverse of loop distribution      \\
%	\joop{reorder}                                          & Generalized distribution, fusion, statement reordering \\
	\joop{wavefront}                                        & Loop skewing                                         \\
	\joop{unrollandjam}                                     & Unroll-and-jam/register tiling                       \\
	\joop{interleave}                                       & Loop interleaving                                    \\
	\joop{scatter}                                          & Polyhedral scheduling                                \\
%	\joop{balance}                                          & Work balancing                                       \\
	\joop{curve}                                            & Space-filling curve (Hilbert-, Z-curve or similar)   \\
	\joop{assume\_coincident}                               & Assume no loop-carried dependencies                  \\
	\joop{assume\_parallel}                                 & Assume parallelism                                   \\
	\joop{assume\_min\_depdist}                             & Assume minimum dependence distance                   \\
	\joop{assume\_unrelated}                                & Assume statements access only disjoint memory        \\
	\joop{assume\_termination}                              & Assume that a loop will eventually terminate         \\
	\joop{expect\_count}                                    & Expect an average number of loop iterations          \\
\caption{Directives on loops}\label{tbl:loopdirectives}
\end{longtable}}

% noassert, assume_safety, noversioning, suggest_only
Most clauses are specific to a transformation (e.g. tile sizes), but some clauses apply to all transformations, such as:
\begin{description}

\item[(\texttt{no})\texttt{assert}] 
Control whether the compiler has to abort with an error if, for any reason, a transformation can not be applied.
The default is \texttt{noassert}, but the compiler may still warn about not applied transformations.
%The warning can be suppressed with a \cinline!nowarn! clause, or the compiler's native mechanism.

\item[\texttt{noversioning}] 
Disable code versioning.
If versioning is required to preserve the loop's semantics, do not apply the transformation unless \texttt{assume\_safety} is used as well.
The combination \texttt{assert~noversioning} can be used to ensure that the transformed code always runs instead of a some fallback version.

\item[\texttt{assume\_safety}] 
Assume that the transformation is semantically correct in all well-defined cases.
This shifts the responsibility of correctness to the programmer.
If the compiler was able to apply the transformation using code versioning, in general, this will remove the runtime checks.

\item[\texttt{suggest\_only}] By default, a transformation pragma overrides any profitability heuristic the compiler might use to apply a transformation.
This clause can be used together with \texttt{assume\_safety} to only imply that the transformation is semantically correct, but leave it to the profitability heuristic to decide whether to actually apply it, with only a bump in favor of applying the transformation and/or its parameters\meinersbur{e.g. SIMD-width, unroll-factor}.
The compiler might also apply a different transformation.
For instance, \texttt{parallel~for~assume\_safety~suggest\_only} implies that a loop is parallelizable, but the compiler might choose to vectorize it instead.

\end{description}

% Loop hints
Directives with \texttt{assume\_}-prefix inform the compiler that a property is always true.
It is the programmer's responsibility to ensure that this the case and executions that violate the assumption have \emph{undefined behavior}.
Directives with an \texttt{expect\_}-prefix suggest that compiler optimize the code assuming that the property likely applies.
Executions that violate the expectation may run slower, but the behavior remains the same.

\begin{comment}
% __builtin_assume, __builtin_expect (range)
Many compilers also support \cinline!__builtin_assume! and \cinline!__builtin_expect! or equivalents.
These might be also used to by the compiler to improve the applicability of a transformation.  For instance, \cref{lst:builtinexpect} should be equivalent to \cref{lst:pragmaexpect}.
However, if the number of iterations only samples around 128, then \cref{lst:pragmaexpect} is more useful.  
Otherwise, the compiler might emit a special case just for $n=128$, which is different from the intended meaning of a non-trivial number of iterations.

\begin{listing}
\begin{subfigure}[t]{0.5\linewidth}
\begin{minted}{c}
__builtin_expect(n==128);
for (int i=0; i<n; i+=1) 
  { ... }
\end{minted}
\caption{Expected number of iterations using builtin}\label{lst:builtinexpect}%
%
\begin{minted}{c}
#pragma expect_count exact(128)
for (int i=0; i<n; i+=1) 
  { ... }
\end{minted}
\caption{Expected number of iterations using pragma}\label{lst:pragmaexpect}
\end{subfigure}%
\begin{subfigure}[t]{0.5\linewidth}
\begin{minted}{c}
#pragma expect_count avg(128)
for (int i=0; i<n; i+=1) 
  { ... }
\end{minted}
\caption{Average number of iterations}\label{lst:pragmaavg}
\end{subfigure}
\caption{Expected loop iteration}\label{lst:iterations}
\end{listing}
\end{comment}

% Directives on sections
Directives in \cref{tbl:codedirectives} apply to sections of code, which might be in, or include, loops.
%These are mostly hints since a section alone cannot be transformed.
For instance, \texttt{assume\_associative} and \texttt{assume\_commutative} inform the compiler that a section behaves like a reduction, so reduction detection can apply to more than a fixed set of operators.

% Directives on data
%Finally, pragmas in \cref{tbl:datadirectives} apply on data.
%That is, like the previous directives they apply on code sections, instead of applying to the operations in it, they specify a property of memory in its scope, represented by a variable pointing to it.

{\small%
\begin{longtable}{ll}
	Directive                                                               & Short description                                 \\ \hline
	\endhead
	\sect{id}                                                               & Assign an unique name                             \\
	\sect{parallel sections}                                                & OpenMP multi-threaded section                     \\
	\sect{target}                                                           & Accelerator offloading                            \\
	\sect{ifconvert}                                                        & If-conversion                                     \\
	\sect{reorder}                                                          & Execute code somewhere else                       \\
	\data{pack}                                                             & Use copy of array slice in scope                  \\
%	\data{offload}                                                          & Use data copy during accelerator execution \\
	\sect{assume\_associative}                                              & Assume a calculation is associative               \\
	\sect{assume\_commutative}                                              & Assume a calculation is commutative               \\
	\sect{assume\_disjoint\_access}                                         & Memory accesses do not alias                      \\
	\sect{assume\_nooverflow}                                               & Assume (unsigned) integer overflow does not occur \\
	\data{assume\_noalias}                                                  & Assume pointer ranges do not alias                \\
	\data{assume\_dereferenceable}                                          & Assume that a pointer range is dereferenceable    \\
	\sect{expect\_dead}                                                     & Expect that code in a branch is not executed      \\
\caption{Directives on code, including loops}\label{tbl:codedirectives}
\end{longtable}}

In the following we present a selected subset of these transformation in more detail.

\subsubsection{Loop Strip-mining/Blocking/Collapse/Interchange}

are \emph{vertical} loop transformations, i.e., transformations on or to perfectly nested loops%
\meinersbur{Horizontal/Vertical is something I came up with, not insisting on keeping it}.

Strip-mining is the decomposition of a loop into an inner loop of constant size (called the \emph{strip}) and an outer loop (which we call the \emph{pit}\meinersbur{again, an invention by me}) executing that inner loop until all loop bodies have been executed.
For instance, the result of \cref{lst:stripminingexample} is the loop in \cref{lst:stripminingresult}.
%If the inner loop is unrolled, we would have an implementation of partial unrolling.

\begin{listing}
\begin{subfigure}[b]{0.5\linewidth}
\begin{scalepar}{0.95}
\begin{minted}{c}
#pragma omp stripmine strip_size(2) \
        pit_id(outer) strip_id(inner)
for (int i=0; i<128; i+=1)
  Statement(i);
\end{minted}
\end{scalepar}
\caption{Strip-mining pragma}\label{lst:stripminingexample}
\end{subfigure}
\begin{subfigure}[b]{0.5\linewidth}
\begin{scalepar}{0.95}
\begin{minted}{c}
#pragma omp id(outer)
for (int pit_i=0; pit_i<128; pit_i+=2)
  #pragma omp id(inner)
  for (int i=pit_i; i<pit_i+2; i+=1)
    Statement(i);
\end{minted}
\end{scalepar}
\caption{Output loop}\label{lst:stripminingresult}
\end{subfigure}
\vspace*{-4ex}
\caption{Strip-mining example}
\end{listing}

% Loop blocking
The difference of loop blocking is that the pit's size is specified in a clause.
However, the case if the iteration count is not known to be a multiple of the strip/block size requires a different handling.

Collapsing is the reverse operation:
Combine two or more vertical  loops into a single loop.
Only the case where the number of iterations does not depend on anything in the outer loop needs to be supported.
The transformation is already available in OpenMP's \texttt{collapse}-clause of the \texttt{for}-pragma.

%\begin{minted}{c}
%#pragma omp loop(2) collapse
%for (int i=0; i<64; i+=1)
%  for (int j=0; j<64; j+=1)
%    Statement(i,j);
%\end{minted}

% Loop interchange
Interchange is permuting the order of perfectly nested (i.e. vertical) loops.
Classically, only two loops are interchange, but we can generalize this to allow any number of loops as long as they are perfectly nested.
In contrast to the previous transformations, interchange may change the execution order of iterations and therefore requires a legality check.

Using these transformations, other transformations can be constructed.
For instance, tiling is a combination of strip-mining and interchange:

\medskip

\noindent\begin{scalepar}{0.95}
\begin{minted}{c}
#pragma omp interchange permutation(outer_i,outer_j,inner_i,inner_j)
#pragma omp stripmine strip_size(4) pit_id(outer_i) strip_id(inner_i)
for (int i=0; i<128; i+=1)
  #pragma stripmine strip_size(4) pit_id(outer_j) strip_id(inner_j)
  for (int j=0; j<128; j+=1)
    Statement(i);
\end{minted}
\end{scalepar}

\medskip

For convenience, we also propose a \texttt{tile}-transformation which is syntactic sugar for this composition.

% After loop rotation, there might be some residual of the first loop iteration that needs to be put in another horizontal loop

\subsubsection{Loop Distribution/Fusion/Reordering}

are \emph{horizontal} transformations, i.e. apply on loops that execute sequentially (and may nest other loops).

Loop distribution splits a loop body into multiple loops that are executed sequentially.
An example was already given in \cref{lst:distributionexample}.
The opposite is loop fusion: Merge two or more loops into a single loop.
The \texttt{reorder}-pragma changes the execution order of loops or statements.

\begin{comment}
These horizontal operations could be generalized to a single directive, as shown in \cref{lst:rebuild}.
However, we suggest to also support the derived transformations as they are more intuitive and better known.

\begin{listing}
\begin{subfigure}{0.5\linewidth}
\begin{minted}{c}
#pragma omp rebuild part(B) part(C,A)
for (int i=0; i<n; i+=1) {
  #pragma omp id(A)
  { StatementA(i); }
  #pragma omp id(B)
  { StatementB(i); }
}
#pragma omp id(C)
for (int k=0; k<n; k+=1)
  StatementC(i);
\end{minted}
\vspace*{-3ex}%
\caption{Loop rebuild pragma}\label{lst:rebuildexample}
\end{subfigure}%
\begin{subfigure}{0.5\linewidth}
\begin{minted}{c}
for (int i=0; i<n; i+=1)
  StatementB(i);
for (int i=0; i<n; i+=1) {
  StatementC(i);
  StatementA(i);
}
\end{minted}
\vspace*{-3ex}%
\caption{Transformed code}\label{lst:rebuildresult}
\end{subfigure}
\vspace*{-2ex}%
\caption{Generalized loop distribution/fusion}\label{lst:rebuild}
\end{listing}
\end{comment}

\subsubsection{Loop Counter Shifting/Scaling/Reversal}

modifies the iteration space that the compiler associates with a loop.
By itself, this does not do anything, but might be required for other transformations, especially loop fusion.
By default, loop fusion would match iterations that have the same numeric value into the same iteration of the output loop.
If different iterations should be executed together, then the iteration space must be changed such that the instances executed together have the same numeric value.

The scaling transformation only allows positive integer factors.
A factor of negative one would reverse the iteration order which accordingly we call loop reversal.
It may change the code's semantics and therefore requires a validity check.

\subsubsection{Index Set Splitting/Peeling/Concatenation}

modify a loop's iteration domain.
They are \emph{horizontal} loop transformations.
Index set splitting creates multiple loops, each responsible for a fraction of the original loop's iteration domain.
For instance, the result of \cref{lst:splitexample} is shown in \cref{lst:splitresult}.
%The syntax also allows to split the combined iteration space of nested loops.

\begin{listing}
\begin{subfigure}[b]{0.5\linewidth}
\begin{minted}{c}
#pragma omp split indices(i > n/2)
for (int i=0; i<n; i+=1)
  Statement(i);
\end{minted}
\vspace*{-3ex}
\caption{Loop split pragma}\label{lst:splitexample}
\end{subfigure}%
\begin{subfigure}[b]{0.5\linewidth}
\begin{minted}{c}
for (int i=0; i<=n/2; i+=1)
  Statement(i);
for (int i=n/2+1; i<n; i+=1)
  Statement(i);
\end{minted}
\vspace*{-3ex}
\caption{Transformed code}\label{lst:splitresult}
\end{subfigure}
\vspace*{-3ex}
\caption{Index set splitting}
\end{listing}

The difference of \emph{loop peeling} is that the split loop is specified in number of iterations at the beginning or end of the original loop.
Therefore, it can be seen as syntactical sugar for index set splitting.

\emph{Loop concatenation} is the inverse operation and combines the iteration space of two or more consecutive loops.
For instance, the result of \cref{lst:concatenateexample} is \cref{lst:concatenateresult}.

\begin{listing}
\begin{subfigure}[b]{0.5\linewidth}
\begin{minted}{c}
#pragma omp loop(A,B) concatenate 
#pragma omp id(A)
for (int i=0; i<n; i+=1)
  StatementA(i);
#pragma omp id(B)
for (int i=0; i<n; i+=1)
  StatementB(i);
\end{minted}
\vspace*{-3ex}
\caption{Loop concatenation pragma}\label{lst:concatenateexample}
\end{subfigure}%
\begin{subfigure}[b]{0.5\linewidth}
\begin{minted}[escapeinside=!!]{c}
for (int i=0; i<n+n; i+=1) {
  if (i < n)
    StatementA(i);
  else !!
    StatementB(i-n);
}
\end{minted}
\vspace*{-3ex}
\caption{Transformed code}\label{lst:concatenateresult}
\end{subfigure}
\vspace*{-3ex}
\caption{Index set concatenation}
\end{listing}

%%%%%%%%%%%%%%%%%%%%%%%%%%%%%%%%%%%%%%%%%%%%%%%%%%%%%%%%%%%%%%%%%%%%%%%%%%%%%%%%%
\subsection{Syntax}\label{sct:syntax}

% Generalized syntax
The generalized loop transformation syntax we propose is as follows:
\begin{flushleft}
\syntax{\#pragma omp} [\syntax{loop(}\placeholder{loopnames}\syntax{)}] \placeholder{transformation} \placeholder{option}\syntax{(}\placeholder{argument}\syntax{)} \placeholder{switch} \dots
\end{flushleft}
The optional \syntax{loop} clause can be used to specify which loop the transformation applies on.
How many loops can be specified depends on the transformation.
If the clause is omitted, the transformation applies on the following horizontal or vertical loops, depending on the transformation.
Alternatively, a number specified to mean the next $n$ vertical or horizontal loops, like OpenMP's \texttt{collapse}-clause.

% What kinds of loops
OpenMP's current \texttt{simd} and \texttt{for} constructs require canonical for-loops, but implementations may lift that restriction to support more kinds of transformable loops, e.g. while-loops.

% Composite directives (#pragma omp parallel for)

%%%%%%%%%%%%%%%%%%%%%%%%%%%%%%%%%%%%%%%%%%%%%%%%%%%%%%%%%%%%%%%%%%%%%%%%%%%%%%%%%
%%%%%%%%%%%%%%%%%%%%%%%%%%%%%%%%%%%%%%%%%%%%%%%%%%%%%%%%%%%%%%%%%%%%%%%%%%%%%%%%%
\section{Implementation}\label{sct:impementation}

To serve as a proof-of-concept, we are working on an implementation in Clang.
Independently from this proposal, we also want to improve Clang/LLVM's loop transformations capabilities to make it useful for optimizing scientific high-performance applications.

%%%%%%%%%%%%%%%%%%%%%%%%%%%%%%%%%%%%%%%%%%%%%%%%%%%%%%%%%%%%%%%%%%%%%%%%%%%%%%%%%
\subsection{Front-End: Clang}

Clang's current gneral loop transformation syntax, also shown in \cref{tbl:proprietarypragmas}, is
\begin{flushleft}
\syntax{\#pragma clang loop} \placeholder{transformation}\syntax{(}\placeholder{option}\syntax{)} \placeholder{transformation}\syntax{(}\placeholder{option}\syntax{)} \dots
\end{flushleft}
and therefore differs from the proposed OpenMP syntax: the first keywords (\syntax{omp} vs. \syntax{clang loop}), but also for the transformations themselves.
%It allows one option per transformation.
Multiple options can be given by using different variants of the same transformation at the same time, which is ambiguous when composing transformations.

Instead, we implement a hybrid of both syntaxes, which is:
\begin{flushleft}
\syntax{\#pragma clang loop}[\syntax{(}\placeholder{loopnames}\syntax{)}] \placeholder{transformation} \placeholder{option}\syntax{(}\placeholder{argument}\syntax{)} \placeholder{switch} \dots
\end{flushleft}
The current syntax still needs to be supported for at least the transformations currently possible with Clang.

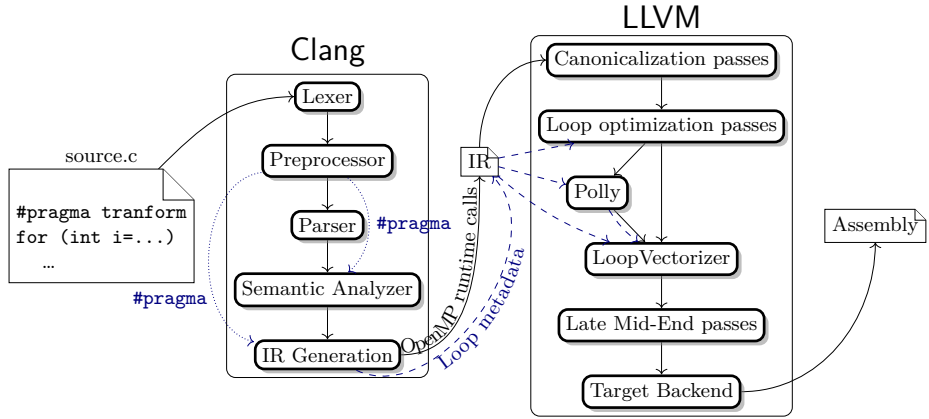
\begin{figure}[t]
\resizebox{\linewidth}{!}{%
\begin{tikzpicture}%
\tikzset{tight/.style={inner sep=0pt,outer sep=0pt,minimum size=0pt}}
\tikzset{node/.style={draw,fill=white,line width=1.2pt,rounded corners,drop shadow}}
\tikzset{supernode/.style={subgraph text none,draw,rounded corners}}
\tikzset{edge/.style={->}}
\graph[layered layout,edges={edge,rounded corners},level sep=5mm,sibling sep=10mm]{
c[as={\minibox{\\\texttt{\#pragma tranform}\\\texttt{for (int i=...)}\\\hspace*{4mm}\dots}},grow=right,draw,shape=file,fill=white,label={source.c}];
ir[as={IR},shape=file,draw,fill=white,nudge=(up:10mm)];
asm[as={Assembly},shape=file,draw,fill=white];

clang [subgraph text none,label={[font=\Large\sffamily]above:Clang}] // [sibling sep=2mm,grow=down,layered layout] {
lexer [as={Lexer},node];
parser [as={Parser},node,grow=down];
preprocessor [as={Preprocessor},node];
sema [as={Semantic Analyzer},node];
codegen [as={IR Generation},node];
lexer->preprocessor->parser->sema->codegen;
};
llvm [subgraph text none,label={[font=\Large\sffamily]above:LLVM}] // [sibling sep=2mm,grow=down,layered layout] {
canonicalization [as={Canonicalization passes},node];
loopopts [as={Loop optimization passes},node];
polly [as={Polly},node];
vectorization [as={LoopVectorizer},node];
latepasses [as={Late Mid-End passes},node];
backend [as={Target Backend},node];
canonicalization->loopopts->vectorization->latepasses->backend;
loopopts->polly->vectorization;
};
c->[in=180]lexer;
codegen->[out=0,in=-90,postaction={decorate,decoration={text along path,text={OpenMP runtime calls},raise=0.3ex}}]ir;
ir->[out=90,in=180]canonicalization;
backend->[out=0,in=-90]asm;
};
\begin{pgfonlayer}{background}
\node[tight,fit={(clang)},supernode]{};
\node[tight,fit={(llvm)},supernode]{};
\end{pgfonlayer}
\path (preprocessor) edge[edge,densely dotted,bend left=50,draw=blue!50!black] node[midway,right,font=\ttfamily,blue!50!black] {\#pragma} (sema);
\path (preprocessor) edge[edge,densely dotted,bend right=80,draw=blue!50!black] node[pos=0.7,left,font=\ttfamily,blue!50!black] {\#pragma} (codegen);
\path (codegen) edge[edge,dashed,bend right=80,draw=blue!50!black,postaction={decorate,decoration={text along path,text={|\color{blue!50!black}|Loop metadata},raise=-1.7ex,pre=moveto,pre length=13mm}}] (ir);
\path (ir) edge[edge,dashed,draw=blue!50!black] (loopopts);
\path (ir) edge[edge,dashed,draw=blue!50!black] (polly);
\path (ir) edge[edge,dashed,draw=blue!50!black,bend right=10] (vectorization);
\path (polly) edge[edge,dashed,draw=blue!50!black,bend right=10] (vectorization);
\end{tikzpicture}%
}\centering
\vspace*{-2ex}%
\caption{Clang compiler pipeline}\label{fig:pipeline}
\end{figure}

Clang's current architecture has two places where loop transformations occur, shown in \cref{fig:pipeline}.
\begin{enumerate}
\item OpenMP's \texttt{parallel~for} is implemented at the front-end level:
	The generated LLVM-IR contains calls to the OpenMP runtime.
\item Compiler-driven optimizations are implemented in the mid-end:
	A set of transformation passes that each consume LLVM-IR with loops and output transformed IR, but metadata attached to loops can influence the passes' decisions.
\end{enumerate}

This split unfortunately means that OpenMP-parallel loops are opaque to the LLVM passes further down the pipeline.
Also, loops that are the result of other transformations (e.g. LoopDistribute) cannot be parallelized this way because it must have happened before.
An exception is, \pragmaomp{simd} which just annotates a loop inside the IR using \texttt{llvm.loop.vectorize.enable} to be processed by the LoopVectorizer pass.
% and parallel access metadata for memory accesses

Multiple groups are working on improving the situation by adding parallel semantics to the IR specification~\cite{tian_sc16,tapir}.
These and other approaches have been presented on LLVM's mailing list~\cite{pir_llvmdev17,pir_llvm18} or its conferences~\cite{pir_llvm16,pir_eurollvm18}.
Until Clang's implementation of OpenMP supports generating parallel IR, we require users to use a different pragma if they want the mid-end to apply thread-parallelism.
In Clang's case, this is \pragma{clang~loop~parallelize\_thread}%
%\footnote{
%We chose \texttt{parallelize\_thread} instead of \texttt{parallelize} because there are multiple forms of parallelism and threading is just one of them.
%The OpenMP-inspired \pragma{clang loop for} does not indicate itself that it is meant to parallelize and assumes it executes in a \pragma{omp parallel} section.
%In our model, the runtime might just start additional worker threads when required.}.

%%%%%%%%%%%%%%%%%%%%%%%%%%%%%%%%%%%%%%%%%%%%%%%%%%%%%%%%%%%%%%%%%%%%%%%%%%%%%%%%%
\subsection{LLVM-IR Metadata}

% Current metadata
Only existing loops can be annotated using the current metadata, but not loops that result from other transformations.
In addition, there is no transformation order and at most one instance of a transformation type and can be specified.
Therefore, a new metadata format is required.

% New metadata
Our changes use loop annotations only to assign loop names.
The transformation themselves are instead a sequence of metadata associated with the function containing the loop.
Each transformation has to lookup the loop it applies on using the result of the previous transformations.

% AutoUpgrade.cpp

% Current loop passes
Current passes that consume the current metadata need to be modified to read the changed format instead.
Due to their fixed order in the pass pipeline however, they can only apply on loops that originate in the source or are the result of passes that execute earlier in the pipeline.

\begin{comment}
\begin{table}[t]
\begin{tabular}{lll}
	LLVM Pass                       & Metadata                              & Pragma in Clang                \\ \hline
	LoopUnroll                      & \texttt{llvm.loop.unroll.*}           & \texttt{clang loop unroll}     \\
	LoopReroll                      & n/a                                   & n/a                            \\
	LoopDistribution                & \texttt{llvm.loop.distribute.*}       & \texttt{clang loop distribute} \\
	LoopVersioningLICM              & \texttt{llvm.loop.licm_versioning.*}  & n/a                            \\
	\multirow{2}{*}{LoopVectorizer} & \texttt{llvm.loop.vectorize.*}        & \texttt{clang loop vectorize}  \\
	                                & \texttt{llvm.loop.interleave.*}       & \texttt{clang loop interleave} \\
	LoopInterchange                 & n/a                                   & n/a                            \\
	(Simple)LoopUnswitch            & n/a                                   & n/a
\end{tabular}
\caption{Loop transformation passes and the pragmas they process}\label{tbl:passes}
\end{table}
\end{comment}

%We can think of a new pass that outputs a warning when there are are loop transformation metadata left in the IR after all loop passes have run.
%This can happen if neither the LLVM pass nor Polly can apply them, or are not enabled in the optimization pipeline (as Polly is not by default).

%%%%%%%%%%%%%%%%%%%%%%%%%%%%%%%%%%%%%%%%%%%%%%%%%%%%%%%%%%%%%%%%%%%%%%%%%%%%%%%%%
\subsection{Loop Transformer: Polly}

Polly~\cite{polly} takes LLVM-IR code and `lifts' is into another representation --\emph{schedule trees}~\cite{verdoolaege14} -- in which loop transformations are easier to express.
To transform loops, only the schedule tree needs to be changed and Polly takes care for the remainder of the work.
%This makes it easier to implement most loop transformations from \cref{tbl:loopdirectives}.

We can implement most transformations from \cref{tbl:loopdirectives} as follows.
First, let Polly create a schedule tree for a loop nest, then iteratively apply each transformation in the metadata to the schedule tree.
For every transformation we can check whether it violates any dependencies and act according to the chosen policy.
When done, Polly generates LLVM-IR from the schedule tree including code versioning.

% Transformational dependencies
If desired, Polly can also apply its loop nest optimizer which utilizes a linear program solver before IR generation. 
We add artificial \emph{transformational dependencies\meinersbur{my invention}} to ensure that user-defined transformations are not overridden.

%%%%%%%%%%%%%%%%%%%%%%%%%%%%%%%%%%%%%%%%%%%%%%%%%%%%%%%%%%%%%%%%%%%%%%%%%%%%%%%%%
%%%%%%%%%%%%%%%%%%%%%%%%%%%%%%%%%%%%%%%%%%%%%%%%%%%%%%%%%%%%%%%%%%%%%%%%%%%%%%%%%
\section{Evaluation}\label{evaluation}

\begin{figure}[t]
\begin{tikzpicture}
\begin{axis}[
xbar,width=105mm,height=48mm,bar width=1.8ex,bar shift=0pt,
ytick=data,symbolic y coords={base,cblas,atlas,openblas,manual,polly,mkl,peak},yticklabels={{-O3 -march=native},{Netlib CBLAS},{ATLAS},{OpenBLAS},{manual},{Polly},{Intel MKL},{theoretical peak}},axis y line=left,enlarge y limits=0.1,           % Y-Axis
xmin=0,xmax=95,xlabel={Execution time (s)},axis x line=bottom,                                                                                              % X-Axis
nodes near coords={\textcolor{black}{\pgfplotspointmeta}},point meta=explicit symbolic,nodes near coords align={horizontal},nodes near coords style={right} %
]

% Just to ensure the labels are there
\addplot[draw=none,fill=none] coordinates {
(0,base) 
(0,cblas)
(0,atlas)
(0,openblas)
(0,manual) 
(0,polly)  
(0,mkl)
(0,peak)   
};

\addplot[draw=black,fill=gray!60] coordinates {
(33.516,cblas)  [33.5s (1.6\%)]
(5.781,atlas)   [5.8s (9\%)]
(4.484,openblas)[4.5s (12\%)]
(0.594,mkl)     [0.59s (89\%)]
};

\addplot[draw=black,fill=blue] coordinates {
(74.891,base)   [74.9s (0.7\%)]
(3.906,manual)  [3.9s (14\%)]
(1.25,polly)    [1.25s (42\%)]
};

\addplot[draw=black,fill=red!60] coordinates {
(0.534,peak)    [0.53s] 
%(0.386,boost)  [0.39s (138\%)]
};

\end{axis}
\end{tikzpicture}
\vspace{-1ex}
\caption{Comparison of matrix-multiplication execution times on an Intel Core i7 7700HQ (Kaby Lake architecture), 2.8 Ghz, Turbo Boost off}\label{fig:dgemm}
\end{figure}
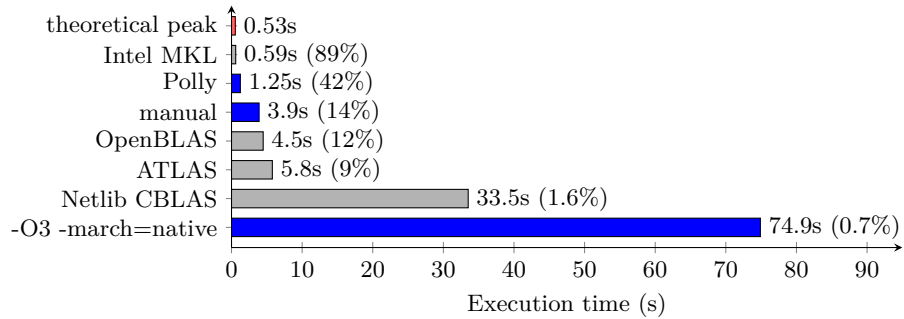

Although we intended this to be a proposal for further discussion, and hence do not have a complete implementation yet, we can measure what the effects of such pragmas are. 
\Cref{fig:dgemm} shows the execution times of a single thread double precision matrix-multiplication kernel ($M=2000$, $N=2300$, $K=2600$).

The na{\"i}ve version (\cref{lst:gemm} without pragmas) compiled with Clang 6.0 executes in 75 seconds (gcc's results are similar); Netlib's reference CBLAS implementation in less than half that time.
With the pragma transformations manually applied the execution time shrinks to 3.9 seconds.
The same transformations as automatically applied by Polly runs in 1.14s, which is 42\% of the processor's theoretical floating-point limit. 
LLVM's loop vectorizer currently only supports vectorizing inner loops, so we applied an additional unroll-and-jam step in the manual version. 
Polly instead prepares its output for the SLP-vectorizer, which may explain the performance difference.

By comparison, OpenBLAS and ATLAS both reach similar results with 4.5 and 5.8 seconds.
Their binaries were obtained from the Ubuntu 16.04 software repository, therefore are likely not optimized for the platform.
Intel's MKL library runs in 0.59 seconds, which is 89\% of the theoretical flop-limited peak performance.

%%%%%%%%%%%%%%%%%%%%%%%%%%%%%%%%%%%%%%%%%%%%%%%%%%%%%%%%%%%%%%%%%%%%%%%%%%%%%%%%%
%%%%%%%%%%%%%%%%%%%%%%%%%%%%%%%%%%%%%%%%%%%%%%%%%%%%%%%%%%%%%%%%%%%%%%%%%%%%%%%%%
\section{Related Work}

% Control over compiler passes
As already mentioned in \cref{tbl:proprietarypragmas}, many compilers already implement pragmas to influence their optimization passes.
The most often implemented transformation is loop unrolling, for instance in gcc since version 8.1~\cite{gccpragmas}.
The most advanced transformation we found is xlc's \pragma{block\_loop}.
It is the only transformation that uses loop names which might have been introduced only for this purpose.
The compiler manual mentions special cases where it is allowed to compose multiple transformations, but in most cases it result in only one transformation being applied or a compiler error~\cite{xlcmanual}.

% Loop transformation composition
Multiple research groups already explored the composition of loop transformations, many of them based on the polyhedral model.  
The Unifying Reordering Framework~\cite{utf} describes loop transformations mathematically, including semantic legality and code generations.
The Clint~\cite{clint} tool is able to visualize multiple loop transformations.

% Source annotation/pragmas/recipes
Many source-to-source compilers can apply the loop transformations themselves and generate a new source file with the transformation baked-in. 
The instructions of which transformations to apply can be in the source file itself like in a comment of the input language (Clay~\cite{clay}, Goofi~\cite{goofi}, Orio~\cite{orio}) or like our proposal as a pragma (X-Language~\cite{xlang}, HMPP~\cite{hmpp}).
Goofi also comes with a graphical tool with a preview of the loop transformations.
The other possibility is to have the transformations in a separate file, as done by URUK~\cite{uruk} and CHiLL~\cite{chill}.
POET~\cite{poet} uses an XML-like description file that only contains the loop body code in the target language.

% Library approach
Halide~\cite{halide} and Tensor Comprehensions~\cite{tensorcomprehensions} are both libraries that include a compiler.
In Halide, a syntax tree is created from C++ expression templates.
In Tensor Comprehensions, the source is passed as a string which is parsed by the library.
Both libraries have objects representing the code and calling its methods transform the represented code.

Similar to the parallel extensions to the C++17~\cite{cpp17} standard library, Intel's Threading Building Blocks~\cite{tbb}, RAJA~\cite{raja} and Kokkos~\cite{kokkos} are template libraries.
The payload code is written using lambdas and an \emph{execution policy} specifies how it should be called.

% Loop autotuning
Our intended use case -- autotuning loop transformations -- has also been explored by POET~\cite{poet} and Orio~\cite{orio}.

%%%%%%%%%%%%%%%%%%%%%%%%%%%%%%%%%%%%%%%%%%%%%%%%%%%%%%%%%%%%%%%%%%%%%%%%%%%%%%%%%
%%%%%%%%%%%%%%%%%%%%%%%%%%%%%%%%%%%%%%%%%%%%%%%%%%%%%%%%%%%%%%%%%%%%%%%%%%%%%%%%%
\section{Conclusion}

We propose adding a framework for general loop transformation to the OpenMP standard.
Part of the proposal are a set of new loop transformations in addition to the already available thread-parallelization (\pragmaomp{for}) and vectorization (\pragmaomp{simd}).
Some of these have already been implemented using compiler-specific syntax and semantics.
The framework allows arbitrarily composing transformation, i.e. apply transformations on already transformed loops.
Loops -- existing in the source code as well as the loop resulting from transformations -- can be assigned unique identifiers such that the pragmas can be applied on already transformed loops.

Experiments show speedups comparable to hand-optimized libraries without the cost in maintainability.
We started implementing the framework using a different syntax in Clang/LLVM using the Polly polyhedral optimizer to carry out the transformations.

The proposal is not complete in that it does not specify every detail a specification would have.
As with any proposal, we are looking for feedback from other groups including about applicability, syntax, available transformations and compatibility/consistency with the current OpenMP standard.

%%%%%%%%%%%%%%%%%%%%%%%%%%%%%%%%%%%%%%%%%%%%%%%%%%%%%%%%%%%%%%%%%%%%%%%%%%%%%%%%%
%%%%%%%%%%%%%%%%%%%%%%%%%%%%%%%%%%%%%%%%%%%%%%%%%%%%%%%%%%%%%%%%%%%%%%%%%%%%%%%%%
\section{Acknowledgments}

This research was supported by the Exascale Computing Project (17-SC-20-SC), a collaborative effort of two U.S. Department of Energy organizations (Office of Science and the National Nuclear Security Administration) responsible for the planning and preparation of a capable exascale ecosystem, including software, applications, hardware, advanced system engineering, and early testbed platforms, in support of the nation’s exascale computing imperative.

This research used resources of the Argonne Leadership Computing Facility, which is a DOE Office of Science User Facility supported under Contract DE-AC02-06CH11357.

%%%%%%%%%%%%%%%%%%%%%%%%%%%%%%%%%%%%%%%%%%%%%%%%%%%%%%%%%%%%%%%%%%%%%%%%%%%%%%%%%
%%%%%%%%%%%%%%%%%%%%%%%%%%%%%%%%%%%%%%%%%%%%%%%%%%%%%%%%%%%%%%%%%%%%%%%%%%%%%%%%%

\bibliographystyle{splncs04}
\bibliography{bibliography}

\end{document}